\documentclass[12pt, letterpaper]{article}

\usepackage{arxiv}
\usepackage[utf8]{inputenc} % allow utf-8 input
\usepackage[T1]{fontenc}    % use 8-bit T1 fonts
\usepackage{amsmath,amsfonts,amssymb,bbm}
\usepackage{mathtools}
\usepackage{mathptmx}
\usepackage{newtxtext}
\usepackage{newtxmath}
\usepackage{hyperref}
\usepackage{caption}
\usepackage{float}
\usepackage{xcolor}
\usepackage{multirow}
\usepackage{subcaption,booktabs}
\usepackage{rotating}
\usepackage{adjustbox}
\usepackage{graphicx}
\usepackage{array}
\usepackage{environ}
\usepackage[authoryear]{natbib}
\title{Single Changepoint Procedures}

\author{
Robert Lund \\
Department of Statistics,\\
University of California, Santa Cruz \\
  \texttt{rolund@ucsc.edu} \\
 \And
Xueheng Shi \\
Department of Statistics,\\
University of Nebraska-Lincoln \\
\texttt{shixueheng@gmail.com}
}

\begin{document}

\maketitle

\begin{abstract}
Single changepoint tests have become a staple check for homogeneity of a climate time series, suggesting how climate has changed should non-homogeneity be declared. This paper summarizes the most prominent single changepoint tests used in today's climate literature, relating them to one and other and unifying their presentations. Asymptotic quantiles for the individual tests are presented. Derivations of the quantiles are given, enabling the reader to tackle cases not considered within.  Our work here studies both mean and trend shifts, covering the most common settings arising in climatology. SOI and global temperature series are analyzed within to illustrate the techniques.
\end{abstract}

\section{Introduction}
\label{sec:introduction}
Structural changes commonly occur in climate time series.  These changes can arise in the series' mean, variance, autocorrelation, or general marginal distributions.  In climatology, structural changes called changepoints are often induced by artificial means such as station relocations and/or instrumentation changes.  Changepoints can also naturally arise \citep{rodionov2005brief}. \cite{Mitchell_1953} estimates that US climate stations average about six relocations or gauge changes per century, about half of which induce mean shift changepoints. Statistical trend analyses are usually distrusted on climate data that has not been adjusted for the effects of changepoints; \cite{Lund_Reeves_2002} and  \cite{reeves2007review} show what can go wrong. In the last two decades, the climate community has recognized the importance of changepoints, and the changepoint homogeneity literature has now exploded.  

Many distinct changepoint methods have been proposed by researchers.  While most of these have merit, others are off base. Unfortunately, changepoint statistics, which are typically based on maximums of dependent quantities, are some of the more technically challenging statistics to quantify; analysis mistakes are easily made. Our intent here is to present an overview of single changepoint analyses in climatology.  We consider changes in mean and/or trend, and link the various statistics to one and other en route.  Our treatment contains most of the settings that arise in practice.  Asymptotic quantiles of the various statistics are presented.

Single changepoint techniques were introduced in \cite{page1955test}, who considered testing for a shift in the mean of a process occurring at an unknown time.  Single changepoint mean shift asymptotic distributions were rigorously quantified in \cite{macneill1974tests}.  Mean shift reviews include \cite{reeves2007review}, \cite{Shi_etal_2021}, \cite{wang2020univariate}, and \cite{lund2023good}.  Staple mean shift tests include cumulative sum (CUSUM) procedures and likelihood ratio tests (LRTs) (also called a standard normal homogeneity test (SNHT)).  Other mean shift references in climatology include \cite{Alexandersson-1986-SNHT}, \cite{ peterson1998homogeneity}, \cite{caussinus2004detection}, and \cite{wang2007penalized} --- this list is necessarily incomplete.

Beyond shifts in process means, changepoints also occur in series with trends. For one example, \cite{beaulieu2024recent} consider whether or not global warming has accelerated recently by examining changes in slopes through changepoint methods.  After our mean shift treatment, this paper moves to changepoint tests for linear trend slopes. Several settings arise here, depending on whether or not one demands continuity of the fitted regression response at the changepoint time.  These cases are enumerated and their analyses are presented.

No climate manuscripts have previously quantified the asymptotic null quantiles of single changepoint methods in any systematic manner. Even statistical books on changepoints such as \cite{Csorgo_1997_LimitThm} focus on specific tests (in this case, likelihood ratios).  Here, an attempt is made to provide reliable null hypothesis quantiles for many of the common single changepoint statistics that the climatologist may encounter.  Our intent is to construct a ``one-stop shopping superstore'' for the climate changepoint community.  While some recommendations on superior and inferior techniques are made en route, comparison has been dealt with elsewhere \citep{Shi_etal_2021, lund2023good} and our focus is more on completeness.

Despite our over-arching intent, some concessions on scope need to be made up front.  First and foremost, we do not delve into multiple changepoints. The tests considered here are at most one changepoint (AMOC) procedures. Here, the null statistical hypothesis is that the series is homogeneous (no changepoints) and is assessed against the alternative hypothesis that one changepoint occurs at an unknown time. While many researchers devise multiple changepoint procedures from AMOC tests via a procedure known as binary segmentation, binary segmentation is arguably the worst way to handle multiple changepoint problems \citep{Shi_etal_2021, lund2023good}.  Second, our quantiles are for Gaussian series only; non-Gaussian marginal distributions can be handled as in \cite{Lund_etal_2024}, but this is beyond our scope here. Finally, we consider independent observations only; mistakes are easily made in changepoint analyses when autocorrelation between observations is ignored \citep{lund2023good}.

The rest of this paper proceeds as follows. Section \ref{sec:meanshift} considers mean shift detection, perhaps the most common changepoint analysis. Section \ref{sec:trendshift} moves to linear trend shifts.  Section \ref{sec:app} presents two examples showing how the tests work and Section \ref{sec:diss} closes with discussion. 

\section{Mean Shift Changepoint Tests}
\label{sec:meanshift}

Perhaps the simplest changepoint test discerns whether or not the series has a single mean shift.  The shift occurs at an unknown time; should the time of the changepoint be known {\em apriori}, the shift time is called a breakpoint and intervention methods are the appropriate analysis tools.

Given a time series $\{X_t\}_{t=1}^{t=n}$, the regression model for a mean shift changepoint occurring at the unknown time $k$ is
\begin{align}
	X_t= \begin{cases}
		\mu + \epsilon_t, \qquad         & \text{for $1   \le t \le k$,}\\
		\mu + \Delta +\epsilon_t, \qquad & \text{for $k+1 \le t \le n$,}
	\end{cases}.
	\label{eqn:means-shift model}
\end{align}
in (\ref{eqn:means-shift model}), $\mu$ is an unknown location parameter, $\Delta$ is the magnitude of the time $k$ mean shift, and $\{ \epsilon_t \}_{t=1}^{t=n}$ is an independent and identically distributed Gaussian white noise having mean zero and variance $\sigma^2$. The null ($H_0$) and alternative ($H_A$) hypotheses for a single changepoint are $H_0: \Delta=0$ versus $H_A: \Delta  \neq 0$. In this section, no trend is assumed to exist. 

\subsection{$Z_{\rm max}$ and CUSUM Statistics}

Suppose that the changepoint was {\em apriori} known to occur at time $k$ and that $\mbox{Var}(X_t) \equiv \sigma^2$ is also known.  Testing whether the two segment means are equal can be accomplished with a $Z$-based statistical test. A two-sample $Z$ test statistic at index $k$ is 
\begin{equation}
	\label{eqn:Z-statistics}
	Z_k=\frac{\overline{X}_{1:k}-\overline{X}_{k+1:n}}
	{ \sqrt{\mbox{Var}(\overline{X}_{1:k}-\overline{X}_{k+1:n})} },
\end{equation}
where $\overline{X}_{1:k}=k^{-1}\sum_{t=1}^k X_t$ and $\overline{X}_{k+1:n}=(n-k)^{-1}\sum_{t=k+1}^n X_t$ denote the sample means of $\{ X_t \}$ before and after time $k$. One rejects mean equality if $|Z_k|$ is too large to be explained by chance variation.

Since $\{ X_t \}$ is Gaussian and we have standardized the statistic by its standard deviation, $Z_k$ has a standard normal distribution (mean zero and unit variance) for each $k$. Simple calculations give
\[
\mbox{Var}(\overline{X}_{1:k}-\overline{X}_{k+1:n})=
\sigma^2 \left[ \frac{1}{k} + \frac{1}{n-k} \right]=
\sigma^2\frac{n}{k(n-k)}.
\]
In particular, one rejects equality of segment means with confidence 95\% if
\[
|Z_k|=\frac{|\overline{X}_{1:k}-\overline{X}_{k+1:n}|}
{\sigma \sqrt{\frac{1}{k} + \frac{1}{n-k}}}
> 1.96.
\]

In practice, $\sigma^2$ is usually unknown.  Its estimated value can be used in place of it (as we do here) without altering any future asymptotic results. This estimate should be calculated under the null hypothesis:
\begin{equation}
	\label{samplevar}
	\hat{\sigma}^2 = \frac{ \sum_{t=1}^n (X_t - \bar{X}_{1:n})^2}{n-1}.
\end{equation}

Cumulative sum (CUSUM) statistics and processes are often used in changepoint analyses.  At index $k$, the CUSUM statistic is
\[
\mbox{CUSUM}_k := \frac{\sum_{t=1}^k X_t - \frac{k}{n} \sum_{t=1}^n X_t}{\sqrt{n}}.
\]

When the changepoint time is unknown, the time where the maximal discrepancy between the two segments occurs is used as the estimate of the changepoint time. Hence, a statistic to detect a mean shift at an unknown time becomes 
\begin{equation}
	\label{eq:Zmaxstat}
	Z_{\max} = \max_{1 \leq k < n}|Z_k| = 
	\max_{1 \leq k < n} 
	\frac{|\mbox{CUSUM}_k|}{\sqrt{ \frac{k}{n}\left(
			1-\frac{k}{n} \right)}}.
\end{equation}
The second equality in (\ref{eq:Zmaxstat}) follows from some elementary algebraic manipulations in the Appendix. Since $Z_k$ accounts for a differing number of observations before and after time $k$, $Z_{\rm max}$ does too.

To assess whether a mean shift changepoint occurs at an undocumented time, one needs to derive the distribution of $Z_{\rm max}$. This is typically done asymptotically and requires advanced probability theory involving weak convergence in function spaces \citep{billingsley2013convergence}. The reader is warned that mistakes have been made in pursuit of this objective.  The ideas are now presented in an elementary manner.  One can quantify the asymptotics in terms of $\{ Z_k \}$ or $\{ \mbox{CUSUM}_k \}$; we choose the latter since this is the traditional statistics route.

From the assumed normality, the Appendix derives the Gaussian relationship
\[
\text{CUSUM}_k \sim \text{N} \left( 0, \sigma^2 \frac{k}{n} \left( 1-\frac{k}{n} \right) \right)
\]
and the covariance
\begin{equation}
	\label{BBmoment}
	\mbox{Cov} \left( \mbox{CUSUM}_{k_1}, \mbox{CUSUM}_{k_2} \right) = \sigma^2
	\frac{k_1}{n} \left( 1-\frac{k_2}{n} \right), \quad 1 \leq k_1 \leq k_2 \leq n.
\end{equation}

This allows us to transition to a Gaussian process in continuous time.  First, if $k$ depends on the sample size $n$ such that $k/n \rightarrow t \in (0,1)$ as $n \rightarrow \infty$, then $\text{CUSUM}_k \overset{{\cal D}}{\rightarrow} N(0,t(1-t))$, where the ${\cal D}$ denotes convergence in distribution as $n \rightarrow \infty$.  For bivariate pairs (CUSUM$_{k_1}$, CUSUM$_{k_2}$) with $k_1/n \rightarrow t \in (0,1)$ and $k_2/n \rightarrow s \in (0,1)$ with $t \leq s$, the covariance in (\ref{BBmoment}) yields 
\[
\left(
\begin{array}{c}
	\mbox{CUSUM}_{k_1} \\
	\mbox{CUSUM}_{k_2} \\
\end{array}
\right)
\stackrel {{\cal D}} {\longrightarrow} \text{N}_2 \left( 
\left(
\begin{array}{c}
	0 \\
	0 \\
\end{array}
\right),
\sigma^2 
\left[ 
\begin{array}{cc}
	t(1-t) & t(1-s) \\
	t(1-s) & s(1-s) \\
\end{array}
\right]
\right), 
\]

From this, some advanced probability theory shows that $\{ B(t) \}_{t=0}^{t=1}$ defined by $B(t) := \lim_{n \rightarrow \infty} \mbox{CUSUM}_k$ converges to a Gaussian process with mean $E[B(t)] \equiv 0$, $\mbox{Var}(B(t))= \sigma^2 t(1-t)$, and $\mbox{Cov}(B(t),B(s))=\sigma^2 t(1-s)$ for $0 \leq t \leq s \leq 1$.  This is the famous Brownian bridge process of empirical statistics \citep{van1996weak}. A standard ($\sigma=1$) Brownian bridge can be constructed from the classical standard Brownian motion process $\{ W(t) \}_{t=0}^{t=1}$ via
\[
B(t)=W(t)-tW(1), \qquad 0 \leq t \leq 1.
\]

We now deal with the maximum and absolute value in (\ref{eq:Zmaxstat}). Because the maximum of the absolute value of a function defined over [0,1] is a continuous operation (over the space of continuous functions), a result called Donsker's Theorem in the probability literature allows us to assert that 
\begin{equation}
	Z_{\max} \stackrel{{\cal D}}{\longrightarrow} 
	\sup_{t \in (0,1)} \frac{|B(t)|}{\sqrt{t(1-t)}}
\end{equation}
as $n \rightarrow \infty$.  There is, however, one important issue.

Technically, the above supremum, taken over $t \in (0,1)$, is infinite; that is, the limiting distribution of $Z_{\rm max}$ is infinite with probability one. This follows from the fact that $\sup_{0 < t < \delta} |W(t)|/\sqrt{t}$ is infinite for any $\delta > 0$, which in turn follows from the law of the iterated logarithm for random walk processes and Brownian motion \citep{billingsley2013convergence, Horvath_Rice_2024}. The mathematics is telling us that when $t$ is close to zero or unity, insufficient observations lie in one of the segments to reliably make conclusions. To proceed, we set a small truncation (cropping) threshold $\delta >0$ and examine 
\[
Z_{{\rm max}, \delta} := \max_{k : k/n \in (\delta, 1-\delta) } |Z_k|.
\]
This ``cropped" $Z_{\rm max}$ statistic is finite with probability one: 
\[
Z_{{\rm max}, \delta} \stackrel {{\cal D}} {\longrightarrow}
\sup_{t \in (\delta,1-\delta)} \frac{|B(t)|}{\sqrt{t(1-t)}}.
\]
Quantiles for $Z_{{\rm max},\delta}$ for $\delta \in \{ 0.01, 0.05, 0.10 \}$ are listed in Table \ref{tab:scpt-critical-values}.

The CUSUM statistic itself can also be used as a mean shift detector.  One simply examines $\max_{1 \leq k <n} |\mbox{CUSUM}_k|$ as the changepoint statistic; the argument maximizing $|\mbox{CUSUM}_k|$ is the estimate of the changepoint time. There is no need to crop CUSUM values at the two boundaries; indeed, as $n \rightarrow \infty$,
\[
\max_{1 \leq k <n} \frac{|\mbox{CUSUM}_k|}{\hat{\sigma}} \stackrel {{\cal D}} {\longrightarrow} \sup_{t \in (0,1)} |B(t)|,
\]
which is finite with probability one.  Table \ref{tab:scpt-critical-values} lists quantiles for $\max_{1 \leq k < n} |\mbox{CUSUM}_k|/\hat{\sigma}$.  
An improvement to the CUSUM test, in the form of superior detection power, sums CUSUM statistics over all times to assess whether a changepoint is present.  This test, dubbed a ``SCUSUM test", uses
\[
\mbox{SCUSUM}=\frac{1}{n} \sum_{k=1}^n \frac{\text{CUSUM}_k^2}{\hat{\sigma}^2}
\]
for the changepoint existence statistic.  Note that CUSUM and SCUSUM are distinct acronyms. The time of the changepoint is still estimated as the location $k$ that maximizes $|\mbox{CUSUM}_k|$. 

The SCUSUM test won the single changepoint comparison competition in \cite{Shi_etal_2021}, was developed in \cite{Kirch_2006_CUSUM}, and has good false detection properties and superior detection power. To quantify its asymptotic distribution, the continuous mapping theorem of probability theory \citep{billingsley2013convergence} shows that 
\begin{equation}
	\label{scusum}
	\text{SCUSUM} \overset{{\cal D}}{\rightarrow} \int_0^1 B^2(t)dt
\end{equation}
as $n \rightarrow \infty$ under the null hypothesis of no changepoints. Equation (\ref{scusum}) (or simulation) can be used to compute quantiles of the test, which are reported in Table \ref{tab:scpt-critical-values}.

\subsection{Likelihood Ratio Tests (LRT) and Standard Normal Homogeneity Tests (SNHT)}

A staple statistical procedure is the LRT, which compares how likely the series is under the null and alternative hypotheses.  Let $L_k$ denote the Gaussian likelihood of $\{ X_t \}_{t=1}^{t=n}$ when a changepoint occurs at time $k$ and $L_{H_0}$ be the Gaussian likelihood of the series when no mean shift exists. The LRT statistic for a changepoint at time $k$ is
\begin{equation}
	\label{Judas}
	\Lambda_k  =  \frac{\sup_{\mu, \sigma^2} L_{H_0}}{\sup_{\mu, \Delta, \sigma^2} L_{H_k}}.
\end{equation}
Here, $\sup_{\mu, \sigma^2} L_{H_0}$ uses the parameters $\mu$ and $\sigma^2$ that maximize the likelihood when $H_0$ is true and $\sup_{\mu, \Delta, \sigma^2} L_{H_k}$ uses the $\mu$, $\Delta$, and $\sigma^2$ that maximize the likelihood when $H_A$ is true and the shift time is known as $k$. Homogeneity is rejected in favor of a changepoint at time $k$ if $\Lambda_k$ is sufficiently small in a statistical sense, the ``smallness threshold" depending on the desired statistical confidence.

Since $k$ is unknown, a LRT rejects homogeneity when $\min_{1 \leq k < n} \Lambda_k$ is small enough. This leads to rejecting homogeneity when 
\[
\ell_{\rm \max} := \max_{1 \leq k < n} -2 \ln (\Lambda_k)
\]
is too large to be explained by chance variation. To compute the LRT statistic explicitly, we need to derive $\sup_{\mu, \Delta, \sigma^2}L_{H_k}$ and $\sup_{\mu, \sigma^2}L_{H_0}$. The Appendix establishes
\[
\sup_{\mu, \Delta, \sigma^2} L_{H_k} = (2\pi)^{-n/2} (\hat{\sigma}_{H_k}^2)^{-\frac{n}{2}} 
\exp(-n/2), \qquad 
\sup_{\mu, \sigma^2} L_{H_0} = (2\pi)^{-n/2} (\hat{\sigma}_{H_0}^2)^{-\frac{n}{2}} \exp(-n/2)
\]
from the normal distribution governing the observations, where 
\[
\hat{\sigma}_{H_k}^2=\frac{1}{n}
\left( \sum_{t=1}^k (X_t-\bar{X}_{1:k})^2 + \sum_{t=k+1}^n (X_t-\bar{X}_{k+1:n})^2\right), \quad
\hat{\sigma}_{H_0}^2=\frac{1}{n}
\sum_{t=1}^k (X_t-\bar{X}_{1:n})^2.
\]
Here, $\hat{\sigma}_{H_k}^2$ and $\hat{\sigma}_{H_0}^2$ are the likelihood estimate of $\sigma^2$ when a mean shift occurs at time $k$ (compare to (\ref{samplevar})).  Using these in (\ref{Judas}) shows that homogeneity is rejected when 
\[
\ell_{\max} = n \max_{1 \leq k < n } 
\ln \left( \frac{\hat{\sigma}^2_{H_0}}{\hat{\sigma}^2_{H_{k}}} \right)
\]
is too large (the small/large direction shifts after taking a negative logarithm) to be explained by chance variation.  The rejection threshold is set according to the desired statistical confidence.

To make conclusions, the asymptotic distribution of $\ell_{\max}$ is needed. \cite{Csorgo_1997_LimitThm} show that under $H_0$, for every $t$, the extreme value probability law
\begin{align}
	\lim_{n \rightarrow \infty} \mathbb{P} \left[ \sqrt{2 \ell_{\max}  \ln (\ln(n)) } \leq t+2\ln(\ln(n))+\frac{1}{2}\ln(\ln(\ln(n)))-\ln(\sqrt{\pi}) \right] = e^{-2e^{-t}}
	\label{eqn:gumbel-dist}
\end{align}
holds. Equation (\ref{eqn:gumbel-dist}) allows us to compute $p$-values for LRTs. A caveat is made: the convergence rate in (\ref{eqn:gumbel-dist}) is quite slow, implying that LRTs require a large sample size to become trustable. 

A related climatological staple test for a mean shift is the SNHT of \cite{Alexandersson-1986-SNHT}. Assuming unit variance data, the SNHT statistic at index $k$ is  
\[
\text{SNHT}_k = k \overline{X}^2_{1:k} + 
(n-k) \overline{X}^2_{k+1:n}
\]
and the test statistic for the existence of a changepoint is $\max_{1 \leq k < n} {\rm SNHT}_k$. 

The Appendix relates the SNHT to the LRT by showing that ${\rm SNHT}_k = C - 2 \ln (\Lambda_k)$ for some constant $C$ (this constant depends on the data). Hence, the SNHT is also a Gaussian LRT and the SNHT and the LRT are equivalent tests when the process variance is unity. SNHT critical values are obtained by Monte Carlo simulations for different sample sizes \citep{Khaliq-2007-SNHT}. However, the assumption of a unit variance is unrealistic in practice and influences critical values; perhaps more problematic, estimating $\sigma^2$ (and then scaling the data by it to make a unit variance series) when changepoints exist can be problematic.

\subsection{Comments}

It would be shortsighted not to mention $U$-statistics.  Single changepoint statistics are generally two-sample comparisons, measuring discrepancies before and after the changepoint time. As such, many changepoint statistics can be written in the form of $U$-statistics. A $U$-statistic for the candidate changepoint time $k$ has the general form 
\[
U_k := \sum_{i=1}^k \sum_{j=k+1}^n h(X_i, X_j), \qquad 1 \leq k < n,
\]
where $h(\cdot,\cdot)$ is a real-valued function called a kernel. Common choices of $h$ are $h(x,y)=x-y$ and $h(x,y)=|x-y|^r \mathbbm{1}(x \geq y) - |x-y|^r \mathbbm{1}(x<y)$ for continuous data ($r>0$ is a user-specified parameter), and $h(x,y) = \mathbbm{1}(x<y) - \frac{1}{2}$ for discrete data. Here, $\mathbbm{1}(\cdot)$ is the zero/one indicator,  $\mathbbm{1}(x<y)=1$ if $x<y$ and zero otherwise. A detailed review of $U$-statistics for changepoint problems can be found in \cite{Dehling-2022-UStatistics}.

A weighted maximum absolute $U$-statistic for a single mean shift has the form 
\begin{equation}
	\label{U2}
	U_{\rm max} := \max_{1 \leq k < n} 
	\frac{\left|\sum_{i=1}^k \sum_{j=k+1}^n h(X_i, X_j)\right|}
	{\left[ \frac{k}{n}\left(1-\frac{k}{n}\right)\right]^\gamma n^{3/2}},  
\end{equation}
where $\gamma \in [0,1/2)$ is a user-set parameter. The no changepoint null hypothesis is rejected when $U_{\rm max}$ is too large to be explained by random chance. As the limit distribution of $U_{\rm max}$ depends on both $h$ and $\gamma$, critical values will also depend on them; hence, quantiles will need to be developed for each case. As such, we will not delve further into $U$-statistics; however, the case where $\gamma=1/2$ and $h(x,y)= x-y$ generally leads to CUSUM statistics (one needs to crop boundaries again).   

To connect LRT and CUSUM statistics, scale and square $\mbox{CUSUM}_k$ via 
\begin{align*}
	\lambda_k = \frac{\text{CUSUM}_k^2}{\frac{k}{n}\left(1-\frac{k}{n}\right)}.
\end{align*}
Since $x^2$ is continuous in $x$, the continuous mapping theorem of probability and our previous arguments with $Z_{\rm{max}, \delta}$ give
\[
\max_{k: k/n \in (\delta, 1-\delta)} \lambda_k =Z_{\rm{max, \delta}}^2
\overset{{\cal D}}{\rightarrow} 
\sup_{t \in (\delta, 1-\delta)} \frac{B^2(t)}{t(1-t)}
\]
as $n \rightarrow \infty$.  \cite{Robbins-2009-Dissertation} shows that, as $n \rightarrow \infty$,
\[
\max_{k: k/n \in (\delta , 1-\delta)} -2 \ln(\Lambda_k) \xrightarrow{\mathcal{D}} 
\sup_{t \in  (\delta, 1-\delta)} \frac{B^2(t)}{t(1-t)}.
\]
Hence, LRTs and squared, scaled CUSUM statistics have the same asymptotic distributions.

Before closing our mean shift treatment, one might seek recommendations given the myriad of test choices. We recommend the SCUSUM test as a reliable high detection power test (See \cite{Shi_etal_2021}). We do not recommend using the LRT or SNHT tests without careful consideration.

\begin{table}[ht]
	\caption{Asymptotic quantiles for mean shift tests.}
	\begin{center}
		\begin{tabular}{ c l l l l l }
			\hline
			Test Statistic & 90\% & 95\% & 97.5\% & 99.0\% & 99.9\%  \\
			\hline
			$Z_{\max, .01}$ & 2.970 &3.225 &3.455 &3.730 &4.331     \\
			$Z_{\max, .05}$ & 2.833 &3.095 &3.331 &3.619 &4.241    \\
			$Z_{\max, .10}$ & 2.736 &3.007 &3.252 &3.548 &4.171    \\
			CUSUM           & 1.224 & 1.358 & 1.480 & 1.628 & 1.949   \\
			SCUSUM          & 0.347 & 0.461 & 0.581 & 0.743 & 1.168   \\
			\hline
		\end{tabular}
		\label{tab:scpt-critical-values}
	\end{center}
\end{table}

\section{Series with Trends}
\label{sec:trendshift}

Climate time series often exhibit trends and changes may occur in these trends.  For example, changes in linear warming rates in global surface air temperature records were investigated in \cite{beaulieu2024recent} to assess global warming's possible acceleration. In a simple linear regression setting, three distinct cases describe possible shifts: a) a mean shift while maintaining a constant trend slope in both regimes, b) changes in both the intercept and trend slope, and c) a joinpoint model that allows the trend slope to shift, but requires the two regression responses to meet at the changepoint time.  All three models have the null hypothesis (no changepoint) form
\[
X_t = \mu + \alpha t + \epsilon_t, \quad t=1,2, \ldots, n,
\label{eqn:trend-model}
\]
where $\alpha$ is the linear trend slope.  The next three subsections address these three cases, respectively. Trend shifts are not as extensively investigated (relative to mean shift cases) in the statistics literature; much of our subsequent discourse is new. While it may be tempting to examine $X_t - X_{t-1}$ to convert a slope change problem into a mean change problem, doing so induces autocorrelation in the process errors, which is beyond the scope of this paper. 

\subsection{A Mean Shift under a Constant Trend}

An alternative hypothesis regression model allowing for a possible change in the regression intercept at time $k$ while maintaining a constant trend slope obeys
\begin{equation}
	\label{eqn:fixed-trend}
	X_t = \mu + \alpha t + \Delta \mathbbm{1}_{[t> k]} + \epsilon_t, \quad t=1, 2, \ldots, n.
\end{equation}
Here, $\Delta$ is the mean shift size, $\mu$ is the location parameter of the regression during the first regime, and $\alpha$ is the linear trend slope, which is assumed the same in both regimes in this setting.

A changepoint test statistic can be developed by mimicking arguments in the previous mean shift section. Under the changepoint free null hypothesis, least squares parameter estimates are
\begin{equation}
	\label{burp}
	\hat{\alpha} = \hat{\alpha}_{1:n} = 
	\frac{12 \sum_{t=1}^n t (X_t-\bar{X}_{1:n})}
	{n(n+1)(n-1)}, 
	\qquad  
	\hat{\mu} = \hat{\mu}_{1:n} = \overline{X}_{1:n} - \hat{\alpha}_{1:n} \bar{t}_{1:n},
\end{equation}
where $\bar{t}_{1:n}=(n+1)/2$ is the average time index in $\{ 1, \ldots, n \}$. Should a changepoint exist at time $k$, estimates of $\mu$ from data before and after time $k$ should be statistically different. Estimates of $\mu$ from the data in $\{ 1, \ldots, k \}$ and $\{ k+1, \ldots , n \}$ are 
\[
\hat{\mu}_{1:k} = \bar{X}_{1:k} - \hat{\alpha}_{1:n} \bar{t}_{1:k}, 
\quad
\hat{\mu}_{k+1:n} = \bar{X}_{k+1:n} - \hat{\alpha}_{1:n} \bar{t}_{k+1:n},
\]
respectively.  The slope $\alpha$ is estimated in (\ref{burp}) from all $n$ data points since it is common to both regimes. The subscripts $1:k$ and $k+1:n$ indicate which segment the parameters are estimated from.

Define the standardized difference between regime estimators of $\mu$ as
\[
D_k = \frac{\hat{\mu}_{k+1:n}-\hat{\mu}_{1:k}}{\mbox{Var}(\hat{\mu}_{k+1:n}-\hat{\mu}_{1:k})^{1/2}}.
\]
Then $D_k$ should be large when a shift occurs at time $k$. Since we do not know $k$, the maximal absolute deviation $D_{\rm max} = \max_{1 \leq k < n} |D_k|$ is used as a test statistic.  The Appendix derives 
\[
\mbox{Var}(\hat{\mu}_{k+1:n}-\hat{\mu}_{1:k})=\left(\frac{1}{n-k}+\frac{1}{k}\right) \sigma^2-\frac{3 n}{(n+1)(n-1)} \sigma^2, 
\]
and
\[
\mbox{Cov}(D_k, D_\ell) =\frac{\frac{n}{(n-k) \ell}-\frac{3 n}{(n+1)(n-1)}}{\sqrt{\frac{1}{n-k}+\frac{1}{k}-\frac{3 n}{(n+1)(n-1)}} \sqrt{\frac{1}{n-\ell}+\frac{1}{\ell}-\frac{3 n}{(n+1)(n-1)}}}, \qquad 1 \leq k \leq \ell \leq n. \\
\]

To conduct asymptotics, let $n \rightarrow \infty$ in a manner that $k/n \rightarrow t$ and $\ell/n \rightarrow s$. Arguing as in the last section provides $D_{\rm max} \stackrel {{\cal D}} \longrightarrow \sup_{t \in (0,1)} |G(t)|$,
where $\{ G(t) \}_{t=0}^{t=1}$ is a Gaussian process with mean $E[ G(t) ] \equiv 0$, $\mbox{Var}(G(t)) \equiv 1$, and 
\[
\mbox{Cov}(G(t), G(s)) =
\frac{t(1-s)-3t(1-t)s(1-s)}{\sqrt{t(1-t)-3t^2(1-t)^2} \sqrt{s(1-s)-3s^2(1-s)^2}},  \quad 0 < t \leq s < 1.
\]

It is an exercise in probability theory to show that 
$\sup_{t \in (0,1)} |G(t)| = \infty$ with probability one (this follows from Slepian's Lemma \citep{slepian1962one} and comparisons to a Brownian bridge process). Because of this, a boundary-truncated version of the test is needed.  Specifically, we use $D_{{\rm max},\delta} := \max_{k: \delta \leq k/n < 1-\delta} |D_k|$, 
as the test statistic, which has the asymptotic behavior
\[
D_{{\rm max},\delta} \stackrel{{\cal D}}{\longrightarrow} \sup_{t \in (\delta, 1-\delta)} |G(t)|
\]
as $n \rightarrow \infty$.  Asymptotic quantiles for $D_{{\rm max}, \delta}$ are shown in Table \ref{tab:scpt-critical-values_trends} for several $\delta$.

A different test that does not need to truncate boundaries can be devised from the CUSUM statistics of the last section. Residuals under the null hypothesis of no changepoints are 
\[
\hat{\epsilon}_t = X_t-(\hat{\mu}_{1:n} + \hat{\alpha}_{1:n} t). 
\quad t=1, 2, \ldots, n.
\]
A test that CUSUMs the residuals $\{ \hat{\epsilon}_t \}$ is $H_{\rm max} := \max_{1 \leq k < n} |\text{CUSUM}_{\hat{\epsilon}, k}|$, 
where
\[
\text{CUSUM}_{\hat{\epsilon},k}=\frac{1}{\hat{\sigma}_\epsilon \sqrt{n}} \left(\sum_{t=1}^k \hat{\epsilon}_t -
\frac{k}{n} \sum_{t=1}^n \hat{\epsilon}_t \right)
=\frac{1}{\sqrt{n}} \sum_{t=1}^k \hat{\epsilon}_t, 
\quad 
\hat{\sigma}_\epsilon^2=
\frac{\sum_{t=1}^n \hat{\epsilon}_t^2}{n-2}
\]
(we have used that $\sum_{t=1}^n \hat{\epsilon}_t=0$).

The asymptotic properties of $H_{\rm max}$ have been quantified in \cite{Gallagher_etal_2013}: $H_{\rm max} \stackrel{{\cal D}}{\longrightarrow} \sup_{t \in (0,1) } |G(t)|$ as $n \rightarrow \infty$, where $\{ G(t) \}_{t=0}^{t=1}$ is a Gaussian process related to a Brownian bridge via
\[
G(t)=B(t)-6 t(1-t) \int_0^1 B(x) dx.
\]

From Gaussian process theory \citep{ross1995stochastic}, it is known that $\{ G(t) \}_{t=0}^{t=1}$ is another Gaussian process with $G(0)=G(1)=0$, mean $E[ G(t) ] \equiv 0$ for all $t \in[0,1]$, and covariance function
\[
\mbox{Cov}(G(t), G(s))=t(1-s)[1-3s(1-t)], \quad 0 < t \leq s < 1 .
\]
Asymptotic critical values for $H_{\rm max}$ are provided in Table \ref{tab:scpt-critical-values_trends}. It is not necessary to truncate boundaries with this test. One generally prefers this test to the above one when a changepoint exist (or be expected to exist) near the beginning or end of the data record.

\subsection{Changes in Both Slope and Intercept}

When both the slope and intercept are allowed to change at time $k$, the regression has the form
\begin{align}
	X_t = 
	\begin{cases}
		\mu_1 + \alpha_1 t + \epsilon_t, & \qquad 1 \leq t \leq k     \\ 
		\mu_2 + \alpha_2 t + \epsilon_t, & \qquad k+1 \le t \leq n    \\
	\end{cases}.
\end{align}
This is also called a two-phase regression model \citep{Lund_Reeves_2002, wang2003comments}.

Because two parameters can change at the changepoint time, the analysis becomes more difficult. An $F$ statistic can be developed as follows. Should $k$ be the changepoint time, an $F$ test (also called a Chow Test \citep{Chow-1960}) statistic is
\begin{equation}
	F_k=\frac{({\rm SSE}_{\rm Red}-{\rm SSE}_{\rm Full})/2}
	{{\rm SSE}_{\rm Full}/(n-4)}.    
\end{equation}

Homogeneity of the two segments is rejected when $F_k$ is too large to be explained by chance variation.  Here, the sum of squared errors are given by
\begin{align*}
	\rm SSE_{\text{Full}} 
	= \sum_{t=1}^k\left(X_t-\hat{\mu}_1-\hat{\alpha}_1 t\right)^2  +\sum_{t=k+1}^n\left(X_t-\hat{\mu}_2-\hat{\alpha}_2 t\right)^2, \quad 
	\rm SSE_{\text{Red}}  =  \sum_{t=1}^n \left(X_t-\hat{\mu}_{\rm Red}-\hat{\alpha}_{\rm Red} t \right)^2,
\end{align*}
where $\hat{\mu}_{\rm Red}$ and $\hat{\alpha}_{\rm Red}$ are estimated under $H_0$:
\[
\hat{\alpha}_{\rm Red} = \frac
{12 \sum_{t=1}^n t (X_t-\bar{X}_{1:n})}{n(n+1)(n-1)}, 
\qquad  
\hat{\mu}_{\rm Red} = \overline{X}_{1:n}-\hat{\alpha}_{\rm Red} \bar{t}_{1:n}.
\]
When a changepoint exists at time $k$, parameters in the two regimes are estimated via
\begin{equation}
	\begin{alignedat}{2}
		\hat{\alpha}_1&=
		\frac{\sum_{t=1}^k t \left(X_t-\overline{X}_{1:k}\right)}{\sum_{t=1}^k\left(t-\bar{t}_{1:k}\right)^2}, \qquad \qquad \hat{\mu}_1\;&=&\bar{X}_{1:k}-\hat{\alpha}_1 \bar{t}_{1:k} \\
		\hat{\alpha}_2&=\frac{\sum_{t=k+1}^n t \left(X_t-\overline{X}_{k+1:n}\right)}{\sum_{t=k+1}^n\left(t-\overline{t}_{k+1:n}\right)^2}, 
		\qquad
		\hat{\mu}_2&=&\bar{X}_{k+1:n}-\hat{\alpha}_2 \bar{t}_{k+1:n}
	\end{alignedat}
	\label{unc}
\end{equation}
(compare to \ref{burp}).  Since the time of the changepoint is unknown, we take $F_{\rm max} := \max_{2 \leq k \leq n-1} F_k$ as the test statistic, rejecting homogeneity when $F_{\rm max}$ is too large to be explained by chance variation. As a trend slope cannot be estimated without at least two data points, the above maximum is taken over $k \in \{ 2, 3, \ldots, n-2, n-1 \}$. 

The asymptotic distribution of $F_{\rm max}$ is much harder to derive than our previous asymptotic distributions. Nonetheless, \cite{robbins2016general} identifies it as
\[
F_{{\rm max}, \delta} := \max_{k: \delta \le k/n \leq 1-\delta} F_k \stackrel {{\cal D}}{\longrightarrow}
\sup_{\delta \leq t \leq 1-\delta} \frac{1}{2}\boldsymbol{\Lambda}(t)^\prime \boldsymbol{\Omega}(t)^{-1} \boldsymbol{\Lambda} (t)
\]
as $n \rightarrow \infty$, where 
\[
\boldsymbol{\Omega}(t)=
\left[
\begin{array}{cc}
	t-4t^2+6t^3-3t^4          & t^2/2 -2t^3 +7t^4/2 -2t^5 \\
	t^2/2 -2t^3 +7t^4/2 -2t^5 & t^3/3-t^4+2t^5-4t^6/3     \\
\end{array}
\right],
\]
and
\[
\boldsymbol{\Lambda}(t)=
\left(
\begin{array}{cc}
	\kappa_1(t) - \kappa_1(1)(4t-3t^2) - \kappa_2(1) (-6t+6t^2) \\
	\kappa_2(t) -\kappa_1(1)(2t^2-2t^3) -\kappa_2(1)(-3t^2+4t^3)\\
\end{array}
\right).
\]
Here, $\kappa_1(t)=W(t)$, $\kappa_2(t)= tW(t)-\int_0^tW(u) du$, and $\{ W(t)\}_{t=0}^{t=1}$ is standard Brownian motion.

While this limiting distribution is cumbersome, application only requires quantiles, which are presented in Table \ref{tab:scpt-critical-values_trends} for several $\delta$. For this test, boundaries must be cropped.

\subsection{Joinpoint Models}
Models where the regression response in the two phases are constrained to meet are called joinpoint models (also called joinpin). Joinpoint models were used in \cite{beaulieu2024recent} to assess whether global warming rates have recently changed. The asymptotics presented here have not been investigated to date in the statistics literature.

One way to write a regression model describing this scenario is
\begin{equation} 
	\label{eqn:joinpin}
	X_t = 
	\left\{ 
	\begin{array}{lc}
		\mu + \alpha t + \epsilon_t,                    & \qquad 1  \le t   \leq k,     \\
		\mu + \alpha t + \beta(t-k)+ \epsilon_t,        & \qquad k+1 \le    t   \leq n. \\ 
	\end{array} \right.
\end{equation}

The estimate of $\beta$ when a changepoint exists at time $k$, denoted by $\hat{\beta}_k$, should be statistically non-zero. Deriving least squares parameter estimators is a non-trivial endeavor, even when $k$ is known. While calculations here become unwieldy, \cite{SL_2025} provide the representation
\[
\hat{\beta}_k=
\frac{(bc-ad)n}{M} V_1 + 
\frac{(dn-ab)n}{M} V_2 + 
\frac{(a^2-cn)n}{M} V_3,
\]
where $M=(bc-ad)^2-(b^2-ne)(a^2-nc)$, the three random components are
\[
V_1=\sum_{t=1}^n X_t, \quad V_2=\sum_{t=1}^n tX_t, \quad V_3=\sum_{t=k+1}^n (t-k)X_t,
\]
and the coefficients are
\[
a=\sum_{t=1}^n t, \quad b=\sum_{t=k+1}^n(t-k), \quad c=\sum_{t=1}^n t^2, \quad d=\sum_{t=k+1}^t (t-k)t, \quad e=\sum_{t=k+1}^t (t-k)^2.
\]
From these, another calculation in \cite{SL_2025} gives
\[
\mbox{Var}(\hat{\beta}_k)=\frac{3}{n^3}\frac{\left(\frac{k}{n}\right)^3 \left(1-\frac{k}{n}\right)^3+o(n^{-1})}{\left(\frac{k}{n}\right)^6 \left(1-\frac{k}{n}\right)^6+o(n^{-1})}.
\]
Here, $o(n^{-1})$ is a sequence of numbers $r_n$ such that $nr_n \rightarrow 0$ as $n \rightarrow \infty.$

If time $k$ is a changepoint, then $J_k:=\hat{\beta}_k/\mbox{Var}(\hat{\beta}_{k})^{1/2}$ should be large in absolute magnitude (statistically non-zero).  Because the time $k$ is unknown, the quantity $J_{\rm max} = \max_{2 \leq k \leq n-2} |J_k|$ is used as the test statistic for the existence of a joinpoint changepoint.

To get the asymptotic distribution of $J_{\rm max}$ under the null hypothesis of no changepoints, \cite{SL_2025} derive the limit
\[
\lim_{n \rightarrow \infty} \mbox{Cov}(J_k, J_\ell) = \frac{1}{2} \frac{(3s- t - 2st)}{s(1-t)} 
\sqrt{\frac{t(1-s)}{s(1-t)}}, \quad 0 \leq t \leq s \leq 1.
\]
Here, $k \leq \ell$ and $n \rightarrow \infty$ in a manner such that $k/n \rightarrow t$ and $\ell/n \rightarrow s$. 

Arguing as before, we have that $J_{\rm max} \stackrel {{\cal D}} {\longrightarrow} \sup_{t \in (0,1)} |G(t)|$, where $\{ G(t) \}_{t=0}^{t=1}$ is a zero-mean unit-variance Gaussian process with 
\[
\mbox{Cov}(G(t), G(s)) = \frac{3s/2 -t/2-st }{s(1-t)}\sqrt{\frac{t(1-s)}{s(1-t)}}, \quad 0 < t \leq s < 1. 
\]
With this statistic, $\sup_{t \in (0,1)} |G(t)|$ is again infinite. Hence, boundaries must again be truncated and $J_{\rm max, \delta} = \max_{k: \delta < k/n < 1-\delta } |J_k|$ is used as the changepoint detection statistic.   This statistic has the limit
\[
J_{\rm max , \delta} \stackrel {{\cal D}} {\longrightarrow} \sup_{t \in (\delta, 1-\delta)} |G(t)|,
\]
which is finite with probability one.  Asymptotic quantiles for $J_{{\rm max}, \delta}$ are reported in Table \ref{tab:scpt-critical-values_trends} for several common truncation values of $\delta$.

\begin{table}[H]
	\caption{Asymptotic quantiles for trend shift tests.}
	\begin{center}
		\begin{tabular}{ c l l l l l }
			\hline
			Test Statistic & 90.0\% & 95.0\% & 97.5\% & 99.0\% & 99.9\%  \\
			\hline
			$D_{\max, .01}$ & 3.224 & 3.463 & 3.679 & 3.935 & 4.403  \\
			$D_{\max, .05}$ & 3.135 & 3.378 & 3.603 & 3.895 & 4.403 \\
			$D_{\max, .10}$ & 3.082 & 3.330 & 3.559 & 3.834 & 4.376 \\
			$H_{\max}$      & 0.830 & 0.900 & 0.962 & 1.041 & 1.360  \\
			$F_{\max, .01}$ & 6.595 & 7.444 & 8.273 & 9.336 & 11.866 \\
			$F_{\max, .05}$ & 6.166 & 7.017 & 7.846 & 8.907 & 11.510 \\
			$F_{\max, .10}$ & 5.856 & 6.715 & 7.536 & 8.606 & 11.169 \\
			$J_{\max, .01}$ & 2.530 &2.795 &3.038 &3.327 &3.964  \\
			$J_{\max, .05}$ & 2.380 &2.658 &2.908 &3.207 &3.852   \\
			$J_{\max, .10}$ &2.285 &2.570 &2.827 &3.132 &3.792    \\
			\hline
		\end{tabular}
		\label{tab:scpt-critical-values_trends}
	\end{center}
\end{table}

\section{Examples}
\label{sec:app}

\subsection{Example: El-Ni\~no/Southern Oscillation}
\begin{figure}[H]
	\centering
	\includegraphics[scale=0.6]{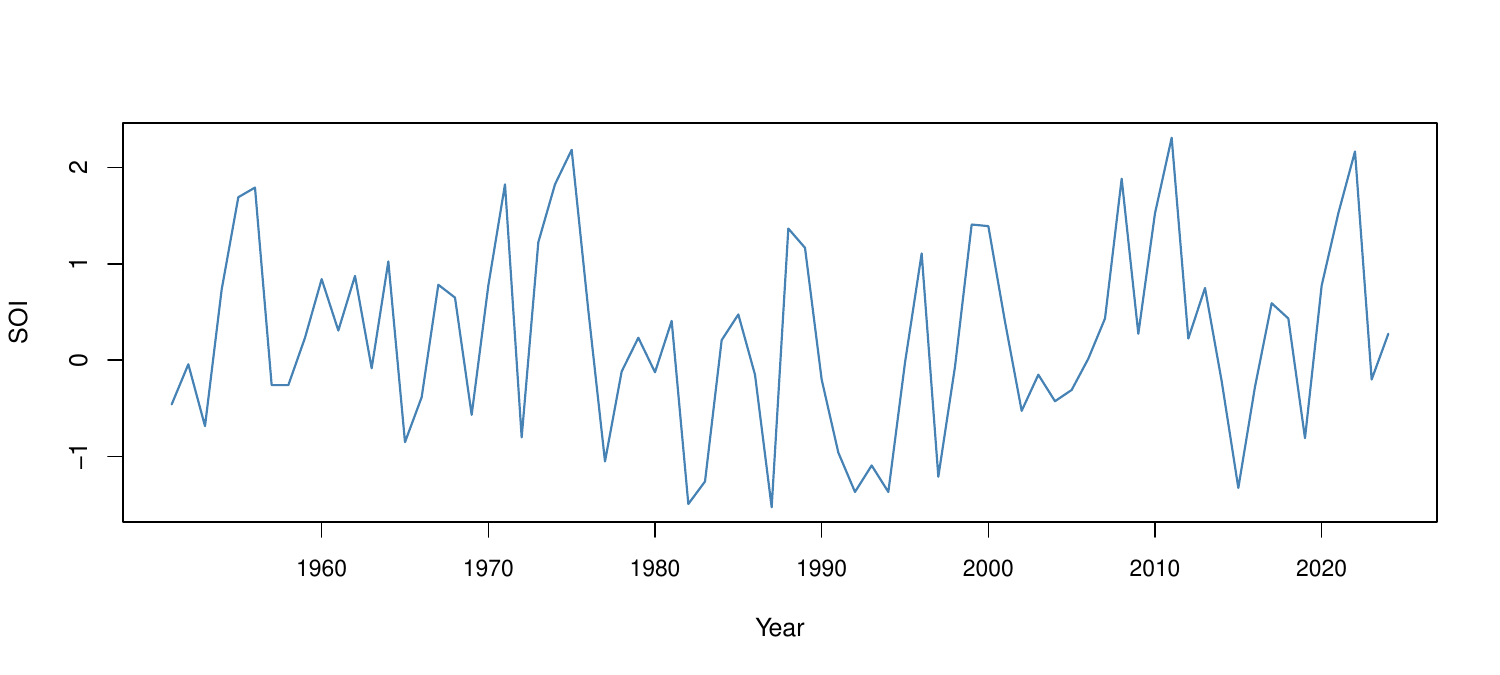}
	\caption{El Ni\~no annual SOI index.}
	\label{fig:elnino3_changepoint}
\end{figure}

The Southern Oscillation Index (SOI) is a standardized index that tracks sea level pressure differences between Tahiti and Darwin, Australia. The SOI is one measure of large-scale fluctuations in air pressure occurring between the western and eastern tropical Pacific (i.e., the state of the Southern Oscillation) during El Ni\~no and La Ni\~na episodes.  The negative phase of the SOI represents below-normal air pressure at Tahiti and above-normal air pressure at Darwin. Our SOI data can be downloaded from \url{https://www.ncei.noaa.gov/access/monitoring/enso/soi}. We scrutinize a 74-year series of annually averaged SOI values for a possible mean shift. Figure \ref{fig:elnino3_changepoint} plots the series from 1951 to 2024.

Table \ref{tab:elnino3_changepoint_summary} reports our Section 2 statistics and their approximate $p$-values as extracted from Table 1 and (\ref{eqn:gumbel-dist})).  All statistics favor the no changepoint null hypothesis with $p$-values exceeding 0.10. The LRT $p$-value was extracted from (\ref{eqn:gumbel-dist}).  Overall, this series appears homogeneous in mean.

\begin{table}[H]
	\centering
	\begin{tabular}{ccccc}
		\hline
		Test & Statistic & Conclusion with confidence $95\%$ & $p$-value bound \\
		\hline
		Zmax ($\delta = 0.01$) & 1.655 & Accept Homogeneity & $p \ge 0.10$ \\
		Zmax ($\delta = 0.05$) & 1.655 & Accept Homogeneity & $p \ge 0.10$ \\
		Zmax ($\delta = 0.10$) & 1.655 & Accept Homogeneity & $p \ge 0.10$ \\
		$\max_k |{\rm CUSUM}_k|$ & 0.783 & Accept Homogeneity & $p \ge 0.10$ \\
		SCUSUM                 & 0.141 & Accept Homogeneity & $p \ge 0.10$ \\
		LRT ($\ell_{\max}$)    & 3.836 & Accept Homogeneity & $p = 0.59$   \\
		\hline
	\end{tabular}
	\caption{Results for mean shift changepoint tests.}
	\label{tab:elnino3_changepoint_summary}
\end{table}

\subsection{Example: Global Temperatures}

\begin{figure}[H]
	\centering
	\includegraphics[width=0.95\textwidth]{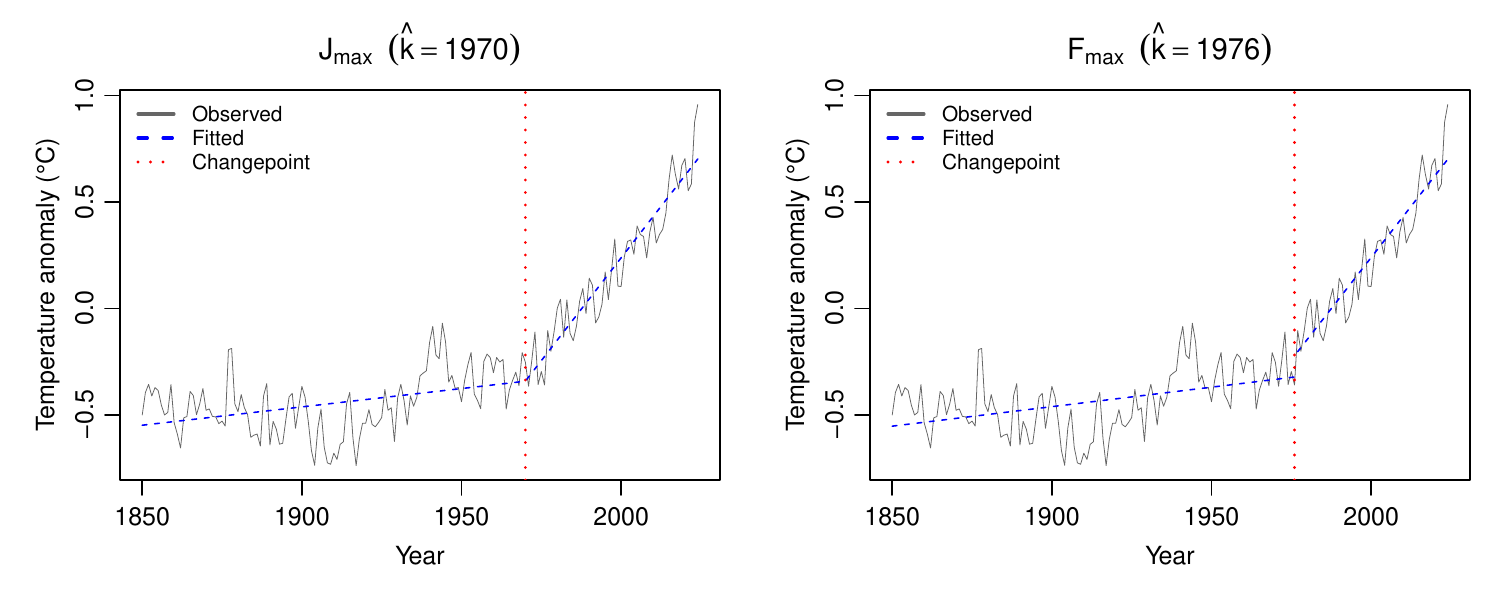}
	\caption{NOAA annual global temperature anomalies during 1850-2023.}
	\label{fig:NOAA_Trend_changepoint}
\end{figure}

Our next application moves to trends, scrutinizing the merged Land–Ocean Surface Temperature record from the National Oceanic and Atmospheric Administration (NOAAGlobalTemp v5.1.0.) of \cite{Vose_al_2021} for shifts.  This series can be downloaded at the website  \url{https://www.ncei.noaa.gov/access/monitoring/climate-at-a-glance/global/time-series/globe/land\_ocean/1/9/1850-2023}. Annual anomalies over the 174-year period 1850-2023 were analyzed for possible slope changes with the methods in Sections 3b) and 3c). Figure \ref{fig:NOAA_Trend_changepoint} plots the data.

Table \ref{tab:amoc_noaa} summarizes our $J_{\rm max}$ and $F_{\rm max}$ slope change tests. Parameters are estimated using 1850 as $t=1$.  Both statistics flag a significant warming rate increase in the 1970s --- roughly a ten-fold increase in slopes --- presumably attributable to anthropogenic global warming. Continuity constrained and discontinuous fits give the same conclusion, although the estimated changepoint time slightly changes. The estimated piecewise means of the series are plotted against the data in Figure \ref{fig:NOAA_Trend_changepoint} and appear to fit the series well.  While Table \ref{tab:amoc_noaa} lists the 95th quantile of the test, $p$-values are extremely small (less than 0.1\% in each fit). We comment that the 1970-2023 series subsegment was further analyzed for a trend shift, but nothing significant was found, agreeing with the findings in \cite{beaulieu2024recent}.

\begin{table}[H]
	\centering
	\caption{Slope change test results with 95\% quantiles. All $p$-values are less than 0.1\%.}
	\label{tab:amoc_noaa}
	\begin{tabular}{lccccccc}
		\toprule
		Test &$\hat{\tau}$ & Test Statistic & $95\%$ Quantile & Left Intercept & Left Slope & Right Intercept & Right Slope \\
		\midrule
		$J_{\rm max}$  & 1970 & 18.759 & 2.658  &  -3.739 & 0.0017 & -38.440 & 0.019 \\
		$F_{\rm max}$  & 1976 &175.346 & 6.835  &  -3.945 & 0.0018 & -38.239 & 0.019 \\
		\bottomrule
	\end{tabular}
\end{table}

\section{Discussion and Comments} 
\label{sec:diss}
This paper justified and unified many of the single changepoint techniques for mean and slope shifts used in today's climate literature. The techniques were related to one and other and a new trend shift test for a joinpoint model was developed. The asymptotic quantiles of the test statistics were reported.

Our treatment is admittedly incomplete. For example, non-parametric changepoint tests were not considered here beyond $U$-statistics. The interested reader is referred to \cite{wolfe1984nonparametric}, \cite{carlstein1988nonparametric}, and \cite{brodsky2013nonparametric} for further information on non-parametric changepoint tests.  Also, only mean and trend shift cases were considered in the paper. Other shift types include changes in series variances/volatilities \citep{lavielle2007adaptive} and marginal distribution shifts \citep{harchaoui2008kernel}.

One common feature of climate series is autocorrelation.  Applying our tests, which were constructed for independent Gaussian series, to autocorrelated series without accounting for autocorrelation, often induces erroneous changepoint conclusions \citep{Robbins_etal_JTSA, Shi_etal_2021, lund2023good}. This caution aside, it is usually easy to modify the techniques presented here to account for autocorrelation by pre-whitening the data with estimates of the autocorrelation. This point is discussed in detail in \cite{Robbins_etal_JTSA} and \cite{lund2023good}. The Gaussian assumption is not as important as that of independence, but cannot altogether be eschewed. The asymptotic arguments presented here apply to non-Gaussian independent data if the test statistic for each changepoint time $k$ converges to a normal distribution (which is usually justifiable via the Central Limit Theorem). This said, we do not advocate applying the methods here to exotic non-Gaussian series such as counts or mixtures of discrete and continuous distributions (the latter would arise with the daily precipitation measurements studied in \cite{Gallagher_etal_2012}).

A few of the asymptotic distributions in this paper involving Brownian bridges admit closed form expressions. For example, 
\[
P \left( \sup_{t \in (0,1)} B(t) > x \right) \quad \mbox{and} \quad 
P \left( \int_0^1 B^2(t) dt > x \right) 
\]
can be determined somewhat explicitly when $\{ B(t) \}_{t=0}^{t=1}$ is a Brownian bridge process \citep{Horvath_Rice_2024,tolmatz2002distribution}. Such expressions are convenient for obtaining exact $p$-values for some changepoint tests. Unfortunately, most of the other changepoint statistics here have asymptotic distributions that are more unwieldy. For these, our tables allow one to bound $p$-values.

%%%%%%%%%%%%%%%%%%%%%%%%%%%%%%%%%%%%%%%%%%%%%%%%%%%%%%%%%%%%%%%%
%% ACKNOWLEDGMENTS 
%%%%%%%%%%%%%%%%%%%%%%%%%%%%%%%%%%%%%%%%%%%%%%%%%%%%%%%%%%%%%%%%
\section*{Acknowledgments} Robert Lund thanks National Science Foundation Grant DMS-2113592 for partial support; Xueheng Shi thanks University of Nebraska-Lincoln Grant ARD-2162251011 for partial support.

%  Keep acknowledgments (note correct spelling: no ``e'' between the ``g'' and
% ``m'') as brief as possible. In general, acknowledge only direct help in
%  writing or research. Financial support (e.g., grant numbers) for the work done, 
%  for an author, or for the laboratory where the work was performed must be 
%  acknowledged here rather than as footnotes to the title or to an author's name.
%  Contribution numbers (if the work has been published by the author's institution 
%  or organization) should be placed in the acknowledgments rather than as 
%  footnotes to the title or to an author's name.

%%%%%%%%%%%%%%%%%%%%%%%%%%%%%%%%%%%%%%%%%%%%%%%%%%%%%%%%%%%%%%%%
% DATA AVAILABILITY STATEMENT
%%%%%%%%%%%%%%%%%%%%%%%%%%%%%%%%%%%%%%%%%%%%%%%%%%%%%%%%%%%%%%%%
\vspace{1cm}

\section{Data Statement}
Webpages where the SOI indices and NOAA temperatures analyzed in this paper can be downloaded were listed where they first appeared. \texttt{R} code is available on \url{https://github.com/shixueheng/AMOC}.

%%%%%%%%%%%%%%%%%%%%%%%%%%%%%%%%%%%%%%%%%%%%%%%%%%%%%%%%%%%%%%%%%%%%%
% APPENDIXES
%%%%%%%%%%%%%%%%%%%%%%%%%%%%%%%%%%%%%%%%%%%%%%%%%%%%%%%%%%%%%%%%%%%%%
%
%% If only one appendix, use

%\appendix

%% If more than one appendix, use \appendix[<letter>], e.g.,

%\appendix[A] 

%% Appendix title is necessary! For appendix title:
\newpage
\section{Appendix}
\subsection{Mean Shift Material}

\subsubsection{CUSUM Asymptotics}
\label{app:cusum}
Under $H_0$, $X_t \overset{IID} {\sim} \textbf{N}( \mu, \sigma^2 )$. Then for each $k$, we have the distributions
\[
\displaystyle \sum_{t=1}^n X_t \sim \textbf{N}(n\mu, n\sigma^2), \qquad \frac{k}{n}\displaystyle\sum_{t=1}^n X_t \sim\textbf{N}\left(k\mu,\frac{k^2}{n}\sigma^2\right),
\qquad 
\displaystyle\sum_{t=1}^kX_t \sim \textbf{N}(k\mu,k\sigma^2).
\]
It now follows that $\sum_{t=1}^kX_t - k/n \sum_{t=1}^nX_t$ has a normal distribution:
\begin{align*}
	\mbox{Var} 
	\left( \sum_{t=1}^kX_t - \frac{k}{n}\displaystyle\sum_{t=1}^nX_t \right) &= 
	\mbox{Var}  \left( \sum_{t=1}^k X_t \right)+
	\mbox{Var}  \left( \frac{k}{n}\sum_{t=1}^n X_t \right) -
	2\text{Cov}\left(\displaystyle\sum_{t=1}^kX_t,\frac{k}{n}\displaystyle\sum_{t=1}^nX_t \right) \\
	&=
	k\sigma^2 + \frac{k^2}{n^2}n \sigma^2 -2\text{Cov}\left(\displaystyle\sum_{t=1}^kX_t,\frac{k}{n}\displaystyle\sum_{t=1}^ kX_t\right) \\
	&=\frac{k(n+k)}{n}\sigma^2-2\frac{k}{n} k \sigma^2 =\frac{k}{n}\left(1-\frac{k}{n}\right)\sigma^2. \\
\end{align*}

This implies that
\[
\text{CUSUM}_k \sim \textbf{N} \left( 0 ,  \frac{k}{n} \left( 1-\frac{k}{n} \right) \sigma^2 \right).
\]
and
\[
\frac{\text{CUSUM}_k}{\sqrt{\frac{k}{n}\left( 1-\frac{k}{n} \right)}} \sim \textbf{N}(0,\sigma^2).
\]

Covariances are 
\begin{align*}
	\mbox{Cov}(\text{CUSUM}_{k_1}, \text{CUSUM}_{k_2})
	&= \mbox{Cov} \left( \frac{1}{\sqrt{n}} \left[ \sum_{t=1}^{k_1}X_t-\frac{k_1}{n}\sum_{t=1}^nX_t\right],\ \frac{1}{\sqrt{n}}\left[\sum_{t=1}^{k_2}X_t-\frac{k_2}{n}\sum_{t=1}^nX_t\right]\right)\\
	&=\frac{1}{n}\mbox{Cov}\left(\sum_{t=1}^{k_1}X_t-\frac{k_1}{n}\sum_{t=1}^n X_t, \sum_{t=1}^{k_2}X_t-\frac{k_2}{n}\sum_{t=1}^nX_t\right)\\
	&=\frac{\sigma^2}{n}\left(k_1-\frac{k_2k_1}{n}-\frac{k_1}{n}k_2 + \frac{k_1}{n}\frac{k_2}{n}n\right)\\
	&=\frac{\sigma^2}{n}\left(k_1-\frac{k_1k_2}{n}\right)\\
	&= \sigma^2 \frac{k_1}{n}
	\left( 1- \frac{k_2}{n} \right), \quad 1 \leq k_1 \leq k_2 \leq n
\end{align*}
as claimed in the paper.

\subsubsection{Connections between $Z_{\rm max}$ and CUSUM statistics}

Our work entails mainly algebraic manipulations.   First,
\begin{align*}
	Z_k
	&= \frac{\bar{X}_{1:k} - \bar{X}_{k+1:n}}{\sqrt{\operatorname{Var}(\bar{X}_{1:k} - \bar{X}_{k+1:n})}}
	= \frac{\frac{1}{k}\sum_{t=1}^{k} X_t - \frac{1}{n-k}\sum_{t=k+1}^{n} X_t }{\sigma \sqrt{\frac{n}{k(n-k)}}} \\
	&= \frac{1}{\sigma} \sqrt{\frac{k(n-k)}{n}}
	\left( \frac{1}{k}\sum_{t=1}^{k} X_t - \frac{1}{n-k}\sum_{t=k+1}^{n} X_t \right) \\
	&= \frac{1}{\sigma} \sqrt{\frac{1}{k(n-k)n}}
	\left( (n-k)\sum_{t=1}^{k} X_t - k\sum_{t=k+1}^{n} X_t \right)  \\
	&= \frac{1}{\sigma} \sqrt{\frac{n}{k(n-k)}}
	\left( \sum_{t=1}^{k} X_t - \frac{k}{n}\sum_{t=1}^{n} X_t \right) \\
	&= \frac{\sum_{t=1}^{k} X_t - \frac{k}{n}\sum_{t=1}^{n} X_t}{\sigma \sqrt{\frac{k(n-k)}{n}}}. \\
\end{align*}

On the other hand,
\[
\frac{\text{CUSUM}_k}{\sqrt{\left(\frac{k}{n}\right)\left(1-\frac{k}{n}\right)}}
= \frac{\sum_{t=1}^{k} X_t - \frac{k}{n}\sum_{t=1}^{n} X_t}
{ \sqrt{n}\sqrt{\left(\frac{k}{n}\right)\left(1-\frac{k}{n}\right)}} \\
= \frac{\sum_{t=1}^{k} X_t - \frac{k}{n}\sum_{t=1}^{n} X_t }
{\sqrt{\frac{k(n-k)}{n}}}.
\]
Therefore, for every $k$,
\[
\sigma Z_k = \frac{\text{CUSUM}_k}
{\sqrt{\left(\frac{k}{n}\right) 
		\left(1-\frac{k}{n}\right)}}
\]
as claimed in the paper.

\subsubsection{LRT Material}

To compute the LRT statistic, we need the maximum likelihood estimators (MLEs) under the null and alternative hypotheses. Under $H_0$, the MLEs are
\[
\hat{\mu} = \overline{X}_{1:n}, \quad \hat{\sigma}^2 = \frac{1}{n} \sum_{t=1}^n (X_t - \bar{X}_{1:n})^2=\hat{\sigma}^2_{H_0}.
\]
Under $H_A$ when the changepoint occurs at time $k$, the MLEs for $\mu$, $\Delta$, and $\sigma^2$ are 
\[
\hat{\mu} = \overline{X}_{1:k}, \quad
\hat{\Delta} = \overline{X}_{k+1:n} - \overline{X}_{1:k}, \quad
\hat{\sigma}^2= \frac{1}{n} \left[ \sum_{t=1}^k (X_t - \bar{X}_{1:k})^2 + \sum_{t=k+1}^n (X_t-\bar{X}_{k+1:n})^2 \right]=\hat{\sigma}^2_{H_k}
\]
Plugging these expressions back into the Gaussian likelihood and simplifying gives
\[
\sup_{\mu, \Delta, \sigma^2} L_{H_k} = 
(2\pi)^{-\frac{n}{2}} 
(\hat{\sigma}_{H_k}^2)^{-n/2} 
\exp\!\left(-\frac{n}{2}\right), 
\quad 
\sup_{\mu, \sigma^2} L_{H_0} = (2\pi)^{-\frac{n}{2}} (\hat{\sigma}_{H_0}^2)^{-n/2} 
\exp\!\left(-\frac{n}{2}\right),
\]

The LRT is hence
\[
\Lambda_k = \frac{\sup_{\mu,\sigma^2} L_{H_0}}{\sup_{\mu,\Delta,\sigma^2} L_{H_k}}=\frac{\hat{\sigma}^2_{H_0}}{\hat{\sigma}^2_{H_k}}.
\]
The expression for $\ell_{\rm \max}=-2\ln(\Lambda_k)$ given in the paper now follows. 

\subsubsection{SNHT Material}
To relate the LRT and the SNHT, take a known $\sigma^2$ and use the MLEs above to get
\begin{align*}
	-2\ln(\Lambda_k)
	& = \sup_{\mu,\Delta}\frac{-1}{\sigma^2}
	\left\{
	\sum_{t=1}^{k}(X_t - \mu)^2 
	+ \sum_{t=k+1}^{n}(X_t - (\mu+\Delta))^2
	\right\}
	- \sup_{\mu}\frac{-1}{\sigma^2}
	\sum_{t=1}^{n}(X_t - \mu)^2 \\
	&=  \frac{1}{\sigma^2}\bigg[
	\underbrace{\sum_{t=1}^n (X_t - \overline{X}_{1:n})^2}_{(1)}
	- \underbrace{\sum_{t=1}^k (X_t - \overline{X}_{1:k})^2}_{(2)}
	- \underbrace{\sum_{t=k+1}^n (X_t - \overline{X}_{k+1:n})^2}_{(3)}
	\bigg]
\end{align*}  

Note that (1) above does not depend on $k$; hence, it can be treated as a constant and dropped in the maximization. The sum of terms (2) and (3) is simplified as
\begin{align*}
	&\sum_{t=1}^k (X_t - \overline{X}_{1:k})^2 + \sum_{t=k+1}^n (X_t - \overline{X}_{k+1:n})^2\\
	&= \sum_{t=1}^k X_t^2 - 2\overline{X}_{1:k}\sum_{t=1}^k X_t + k\overline{X}_{1:k}^2 
	+ \sum_{t=k+1}^n X_t^2 - 2\overline{X}_{k+1:n}\sum_{t=k+1}^n X_t + (n-k)\overline{X}_{k+1:n}^2\\
	&= \sum_{t=1}^n X_t^2 - k\overline{X}_{1:k}^2 - (n-k)\overline{X}_{k+1:n}^2\\
	&= \underbrace{\sum_{t=1}^n X_t^2}_{(4)} - k\overline{X}_{1:k}^2 - (n-k)\overline{X}_{k+1:n}^2.
\end{align*}

Since the sum in (4) does not depend on $k$, it can also be omitted from the maximization.  Hence, the SNHT statistic can be expressed as
\[
\text{SNHT}_k = k\overline{X}_{1:k}^2 + (n-k)\overline{X}_{k+1:n}^2 = C-2\ln(\Lambda_k),
\]
where $C$ is some constant that depends on $X_1, \ldots, X_n$. 

\subsection{Series with Trends}

To verify the calculations in subsection 3.a), we first note that
\begin{align*}
	\hat{\mu}_{k+1:n}-\hat{\mu}_{1:k}
	&=\left(\bar{X}_{k+1:n}-\hat{\alpha}\,\bar{t}_{k+1:n}\right)-\left(\bar{X}_{1:k}-\hat{\alpha}\,\bar{t}_{1:k}\right) \\
	&=\left(\bar{X}_{k+1:n}-\bar{X}_{1:k}\right)-\hat{\alpha}\left(\bar{t}_{k+1:n}-\bar{t}_{1:k}\right).
\end{align*}
Hence,
\begin{align*}
	\text{Var}\left(\hat{\mu}_{k+1:n}-\hat{\mu}_{1:k}\right)
	&=\underbrace{\mathrm{Var}\!\left(\bar{X}_{k+1:n}-\bar{X}_{1:k}\right)}_{(1)}
	+\underbrace{\mathrm{Var}(\hat{\alpha})\left(\bar{t}_{k+1:n}-\bar{t}_{1:k}\right)^2}_{(2)}\\
	&\quad -2\,\underbrace{\mathrm{Cov}\!\left(\bar{X}_{k+1:n}-\bar{X}_{1:k},\,\hat{\alpha}\left(\bar{t}_{k+1:n}-\bar{t}_{1:k}\right)\right)}_{(3)}.
\end{align*}

\paragraph{Term (1).}
Now note that
\[
\bar{X}_{k+1:n}-\bar{X}_{1:k}
=\alpha\Big( \bar{t}_{k+1:n}-\bar{t}_{1:k}\Big)+\Big(\bar{\varepsilon}_{k+1:n}-\bar{\varepsilon}_{1:k}\Big).
\]
Since $\alpha ( \bar{t}_{k+1:n}-\bar{t}_{1:k} )$ is constant, we have
\begin{align*}
	(1)=\mathrm{Var}\left(\bar{\varepsilon}_{k+1:n}-\bar{\varepsilon}_{1:k}\right)
	&=\mathrm{Var}\left(\frac{1}{n-k}\sum_{t=k+1}^n \varepsilon_t-\frac{1}{k}\sum_{t=1}^k \varepsilon_t\right) \\
	&=\frac{\sigma^2}{n-k}+\frac{\sigma^2}{k}.
\end{align*}

\paragraph{Term (2).}
Using $\bar{t}_{k+1:n}-\bar{t}_{1:k}=n/2$ gives $(2)=\frac{n^2}{4} \mathrm{Var}(\hat{\alpha})$.  To get $\mathrm{Var}(\hat{\alpha})$, use
\[
\hat{\alpha}=\frac{12 \sum_{t=1}^n t (X_t-\bar{X}_{1:n})}{n(n+1)(n-1)}
\]
and examine the numerator:
\begin{align*}
	\mathrm{Var} \left(\sum_{t=1}^n t\left(X_t-\bar{X}_{1:n}\right)\right)
	&=\mathrm{Var} \left(\sum_{t=1}^n t\varepsilon_t-\bar{\varepsilon}_{1:n}\sum_{t=1}^n t\right) \\
	&=\underbrace{\mathrm{Var}\!\left(\sum_{t=1}^n t\varepsilon_t\right)}_{(2a)}
	+\underbrace{\mathrm{Var} \left(\bar{\varepsilon}_{1:n}\sum_{t=1}^n t\right)}_{(2b)}
	-2\,\underbrace{\mathrm{Cov}\!\left(\sum_{t=1}^n t\varepsilon_t,\bar{\varepsilon}_{1:n}\sum_{t=1}^n t\right)}_{(2c)}.
\end{align*}
Compute the terms gives
\begin{align*}
	(2a)&=\sum_{t=1}^n t^2 \sigma^2=\frac{n(n+1)(2n+1)}{6} \sigma^2,\\
	(2b)&=\Big(\tfrac{n(n+1)}{2}\Big)^2 \mathrm{Var}(\bar{\varepsilon}_{1:n})
	=\frac{n^2(n+1)^2}{4} \frac{\sigma^2}{n}=\frac{n(n+1)^2}{4}\,\sigma^2,\\
	(2c)&=\Big(\tfrac{n(n+1)}{2}\Big)\mathrm{Cov} \left(\sum_{t=1}^n t\varepsilon_t,\bar{\varepsilon}_{1:n}\right)
	=\frac{n(n+1)}{2}\sum_{t=1}^n \frac{t}{n}\sigma^2
	=\frac{(n+1)n}{2} \frac{n(n+1)}{2} \frac{\sigma^2}{n}
	=\frac{n(n+1)^2}{4}\sigma^2.
\end{align*}
Hence,
\begin{align*}
	\mathrm{Var}\left( \sum_{t=1}^n t\left(X_t-\bar{X}_{1:n}\right) \right)
	&=\frac{n(n+1)(2n+1)}{6}\sigma^2+\frac{n(n+1)^2}{4}\sigma^2-2 \frac{n(n+1)^2}{4}\sigma^2\\
	&=\frac{n(n+1)(2n+1)-\tfrac{3}{2}n(n+1)^2}{6}\sigma^2.    
\end{align*}

Therefore,
\begin{align*}
	\mathrm{Var}(\hat{\alpha})
	&=\Big(\frac{12}{n(n+1)(n-1)}\Big)^2 \mathrm{Var}\!\left(\sum_{t=1}^n t(X_t-\bar{X}_{1:n})\right)
	=\frac{12}{n(n+1)(n-1)} \sigma^2,\\
	(2)&=\frac{n^2}{4} \frac{12}{n(n+1)(n-1)} \sigma^2
	=\frac{3n}{(n+1)(n-1)} \sigma^2.
\end{align*}

\paragraph{Term (3).}
Using $\bar{t}_{k+1:n}-\bar{t}_{1:k}=n/2$, 
\begin{align*}
	(3)&=\mathrm{Cov}\left(\bar{X}_{k+1:n}-\bar{X}_{1:k}, \hat{\alpha} \frac{n}{2} \right)
	=\frac{n}{2} \mathrm{Cov} \left(\bar{\varepsilon}_{k+1:n}-\bar{\varepsilon}_{1:k}, \hat{\alpha}\right).
\end{align*}
A direct calculation akin to those that produced $(2a)$–$(2c)$ and using $\mathrm{Cov}(t\varepsilon_t,\bar{\varepsilon}_{1:n})=t \sigma^2/n$ yields $(3)=\frac{3n}{(n+1)(n-1)} \sigma^2$.  Putting the above three terms together now produces
\begin{align*}
	\text{Var}\!\left(\hat{\mu}_{k+1:n}-\hat{\mu}_{1:k}\right)
	&=\Big(\frac{1}{n-k}+\frac{1}{k}\Big)\sigma^2
	+\frac{3n}{(n+1)(n-1)}\sigma^2
	-2 \frac{3n}{(n+1)(n-1)}\sigma^2 \nonumber\\
	&=\Big(\frac{1}{n-k}+\frac{1}{k}\Big)\sigma^2-\frac{3n}{(n+1)(n-1)}\sigma^2,
\end{align*}
which is as quoted in the paper.

To get $\mbox{Cov}(D_k, D_\ell)$, let $1 \leq k \leq  \ell \leq n$ and define
\[
D_k=\frac{\hat{\mu}_{k+1:n}-\hat{\mu}_{1:k}}{\sqrt{\mathrm{Var}(\hat{\mu}_{k+1:n}-\hat{\mu}_{1:k})}}\!,
\qquad
D_\ell=\frac{\hat{\mu}_{\ell+1:n}-\hat{\mu}_{1:\ell}}{\sqrt{\mathrm{Var}(\hat{\mu}_{\ell+1:n}-\hat{\mu}_{1:\ell})}}.
\]
Then
\[
\mathrm{Cov}(D_k,D_\ell)
=\frac{\mathrm{Cov}\!\left(\hat{\mu}_{k+1:n}-\hat{\mu}_{1:k},\,\hat{\mu}_{\ell+1:n}-\hat{\mu}_{1:\ell}\right)}
{\sqrt{\mathrm{Var}(\hat{\mu}_{k+1:n}-\hat{\mu}_{1:k})\,\mathrm{Var}(\hat{\mu}_{\ell+1:n}-\hat{\mu}_{1:\ell})}}.
\]

Write the numerator as
\begin{align*}
	\mathrm{Cov}\!\left(\hat{\mu}_{k+1:n}-\hat{\mu}_{1:k},\,\hat{\mu}_{\ell+1:n}-\hat{\mu}_{1:\ell}\right)
	&=\underbrace{\mathrm{Cov}(\hat{\mu}_{k+1:n},\hat{\mu}_{\ell+1:n})}_{(1)}
	-\underbrace{\mathrm{Cov}(\hat{\mu}_{k+1:n},\hat{\mu}_{1:\ell})}_{(2)}
	-\underbrace{\mathrm{Cov}(\hat{\mu}_{1:k},\hat{\mu}_{\ell+1:n})}_{(3)}\\
	&\quad +\underbrace{\mathrm{Cov}(\hat{\mu}_{1:k},\hat{\mu}_{1:\ell})}_{(4)}.
\end{align*}

\paragraph{Component (1).}
\begin{align*}
	(1)&=\mathrm{Cov}\!\left(\bar{X}_{k+1:n}-\hat{\alpha}\,\bar{t}_{k+1:n},\;\bar{X}_{\ell+1:n}-\hat{\alpha}\,\bar{t}_{\ell+1:n}\right)\\
	&=\underbrace{\mathrm{Cov}(\bar{X}_{k+1:n},\bar{X}_{\ell+1:n})}_{(1a)}
	-\underbrace{\mathrm{Cov}(\bar{X}_{k+1:n},\hat{\alpha})\,\bar{t}_{\ell+1:n}}_{(1b)}
	-\underbrace{\mathrm{Cov}(\hat{\alpha},\bar{X}_{\ell+1:n})\,\bar{t}_{k+1:n}}_{(1c)}
	+\underbrace{\mathrm{Var}(\hat{\alpha})\,\bar{t}_{k+1:n}\bar{t}_{\ell+1:n}}_{(1d)}.
\end{align*}
More calculations give
\begin{align*}
	(1a)&=\frac{\sigma^2}{n-k},\\
	(1b)&=\frac{3(n+\ell+1)\,k}{n(n+1)(n-1)}\,\sigma^2,\\
	(1c)&=\frac{3(n+k+1)\,\ell}{n(n+1)(n-1)}\,\sigma^2,\\
	(1d)&=\frac{3(n+k+1)(n+\ell+1)}{n(n+1)(n-1)}\,\sigma^2,\\
	\Rightarrow\quad
	(1)&=\left[
	\frac{1}{n-k}
	-\frac{3(n+\ell+1)k}{n(n+1)(n-1)}
	-\frac{3(n+k+1)\ell}{n(n+1)(n-1)}
	+\frac{3(n+k+1)(n+\ell+1)}{n(n+1)(n-1)}
	\right]\sigma^2.
\end{align*}

\paragraph{Component (2).}
\begin{align*}
	(2)&=\mathrm{Cov}\!\left(\bar{X}_{k+1:n}-\hat{\alpha}\,\bar{t}_{k+1:n},\;\bar{X}_{1:\ell}-\hat{\alpha}\,\bar{t}_{1:\ell}\right)\\
	&=\underbrace{\mathrm{Cov}(\bar{X}_{k+1:n},\bar{X}_{1:\ell})}_{(2a)}
	-\underbrace{\mathrm{Cov}(\bar{X}_{k+1:n},\hat{\alpha})\,\bar{t}_{1:\ell}}_{(2b)}
	-\underbrace{\mathrm{Cov}(\hat{\alpha},\bar{X}_{1:\ell})\,\bar{t}_{k+1:n}}_{(2c)}
	+\underbrace{\mathrm{Var}(\hat{\alpha})\,\bar{t}_{k+1:n}\bar{t}_{1:\ell}}_{(2d)}.
\end{align*}
With $k \leq \ell$,
\begin{align*}
	(2a)&=\frac{\ell-k}{(n-k)\ell}\,\sigma^2,\qquad
	(2b)=\frac{3k(\ell+1)}{n(n+1)(n-1)}\,\sigma^2,\\
	(2c)&=-\frac{3(n+k+1)(n-\ell)}{n(n+1)(n-1)}\,\sigma^2,\qquad
	(2d)=\frac{3(n+k+1)(\ell+1)}{n(n+1)(n-1)}\,\sigma^2,\\
	\Rightarrow\quad
	(2)&=\left[
	\frac{\ell-k}{(n-k)\ell}
	-\frac{3k(\ell+1)}{n(n+1)(n-1)}
	+\frac{3(n+k+1)(n-\ell)}{n(n+1)(n-1)}
	+\frac{3(n+k+1)(\ell+1)}{n(n+1)(n-1)}
	\right]\sigma^2.
\end{align*}

\paragraph{Component (3).}
\begin{align*}
	(3)&=\mathrm{Cov}\!\left(\bar{X}_{1:k}-\hat{\alpha}\,\bar{t}_{1:k},\;\bar{X}_{\ell+1:n}-\hat{\alpha}\,\bar{t}_{\ell+1:n}\right)\\
	&=\underbrace{\mathrm{Cov}(\bar{X}_{1:k},\bar{X}_{\ell+1:n})}_{(3a)}
	-\underbrace{\mathrm{Cov}(\bar{X}_{1:k},\hat{\alpha})\,\bar{t}_{\ell+1:n}}_{(3b)}
	-\underbrace{\mathrm{Cov}(\hat{\alpha},\bar{X}_{\ell+1:n})\,\bar{t}_{1:k}}_{(3c)}
	+\underbrace{\mathrm{Var}(\hat{\alpha})\,\bar{t}_{1:k}\bar{t}_{\ell+1:n}}_{(3d)}.
\end{align*}
Here,
\begin{align*}
	(3a)&=0,\qquad
	(3b)=-\frac{3(n+\ell+1)(n-k)}{n(n+1)(n-1)}\,\sigma^2,\\
	(3c)&=\frac{3(\,k+1)\,\ell}{n(n+1)(n-1)}\,\sigma^2,\qquad
	(3d)=\frac{3(\,k+1)(n+\ell+1)}{n(n+1)(n-1)}\,\sigma^2,\\
	\Rightarrow\quad
	(3)&=\left[
	\frac{3(n+\ell+1)(n-k)}{n(n+1)(n-1)}
	-\frac{3(\,k+1)\,\ell}{n(n+1)(n-1)}
	+\frac{3(\,k+1)(n+\ell+1)}{n(n+1)(n-1)}
	\right]\sigma^2.
\end{align*}

\paragraph{Component (4).}
\begin{align*}
	(4)&=\mathrm{Cov}\!\left(\bar{X}_{1:k}-\hat{\alpha}\,\bar{t}_{1:k},\;\bar{X}_{1:\ell}-\hat{\alpha}\,\bar{t}_{1:\ell}\right)\\
	&=\underbrace{\mathrm{Cov}(\bar{X}_{1:k},\bar{X}_{1:\ell})}_{(4a)}
	-\underbrace{\mathrm{Cov}(\bar{X}_{1:k},\hat{\alpha})\,\bar{t}_{1:\ell}}_{(4b)}
	-\underbrace{\mathrm{Cov}(\hat{\alpha},\bar{X}_{1:\ell})\,\bar{t}_{1:k}}_{(4c)}
	+\underbrace{\mathrm{Var}(\hat{\alpha})\,\bar{t}_{1:k}\bar{t}_{1:\ell}}_{(4d)}.
\end{align*}
Here,
\begin{align*}
	(4a)&=\frac{\sigma^2}{\ell},\qquad
	(4b)=-\frac{3(n-k)(\ell+1)}{n(n+1)(n-1)}\,\sigma^2,\\
	(4c)&=-\frac{3(n-\ell)(k+1)}{n(n+1)(n-1)}\,\sigma^2,\qquad
	(4d)=\frac{3(\,k+1)(\ell+1)}{n(n+1)(n-1)} \, \sigma^2, \\
	\Rightarrow\quad
	(4)&=\left[
	\frac{1}{\ell}
	+\frac{3(n-k)(\ell+1)}{n(n+1)(n-1)}
	+\frac{3(n-\ell)(k+1)}{n(n+1)(n-1)}
	+\frac{3(\,k+1)(\ell+1)}{n(n+1)(n-1)}
	\right]\sigma^2.
\end{align*}

Collecting the four components, we get
\begin{align*}
	(1)-(2)-(3)+(4)
	&=\left[\frac{n}{(n-k)\,\ell}-\frac{3n}{(n+1)(n-1)}\right]\sigma^2.
\end{align*}
Plugging in the above variance expressions, we arrive at
\begin{align*}
	\mathrm{Cov}(D_k,D_\ell)
	&=\frac{\dfrac{n}{(n-k)\,\ell}-\dfrac{3n}{(n+1)(n-1)}}
	{\sqrt{\dfrac{1}{n-k}+\dfrac{1}{k}-\dfrac{3n}{(n+1)(n-1)}}
		\;\sqrt{\dfrac{1}{n-\ell}+\dfrac{1}{\ell}-\dfrac{3n}{(n+1)(n-1)}}}
\end{align*}
as claimed in the paper.

%\appendixtitle{Title of Appendix}

%%% Appendix section numbering (note, skip \section and begin with \subsection)
%
% \subsection{First primary heading}

% \subsubsection{First secondary heading}

% \paragraph{First tertiary heading}

%%%%%%%%%%%%%%%%%%%%%%%%%%%%%%%%%%%%%%%%%%%%%%%%%%%%%%%%%%%%%%%%%%%%%
% REFERENCES
%%%%%%%%%%%%%%%%%%%%%%%%%%%%%%%%%%%%%%%%%%%%%%%%%%%%%%%%%%%%%%%%%%%%%
% Make your BibTeX bibliography by using these commands:
\vspace{1cm}

\newpage

\bibliographystyle{plainnat}
\bibliography{arxiv.bib}

\end{document}